# FAINT FIELD GALAXIES:
# A POPULATION OF INTRINSICALLY FAINT OBJECTS
# AT LOW REDSHIFT?

ANA CAMPOS*

*Physics Department, University of Durham*

*Durham DH1 3LE, UK*

## ABSTRACT

To distinguish between the different models proposed to understand the excess of faint field counts over the predictions from non-evolving, a number of redshift surveys have been undertaken. The answer has not arrived yet due to the high rate of incompleteness achieved. Un-identified galaxies have been shown to be bluer than identified ones. Here it is argued that some of these galaxies could be "quiescent" dwarfs, still quite blue but with featureless spectra.

## 1. Introduction

Counting galaxies is one of the classical observational methods in cosmology to measure the value of the density parameter $\Omega_0$, through the study of the geometry of the Universe. The observed number of galaxies per interval of magnitude and unit area is compared with the predictions from different cosmological models. In the simplest case, a basic assumption is made: namely that the luminosity and the number density of galaxies remain both constant on time. Then, given the luminosity function (LF) is straightforward to compute the number count predictions by accounting for the redshift-dependence of the volume elements in the different geometries considered. These basic models are usually called non-evolving.

It has already been shown[1] that non-evolving models fail to match the counts, no matter the value of $q_0$ considered. There is an *excess* of galaxies at faint levels with respect to the model predictions, which is larger when the counts are made using blue photometric bands[2,3]. The excess depends on the normalization of the models. If these are normalized to match the counts at the bright end ($B = 16.5$), the excess is as large as a factor of $\sim 4$ already at $B \sim 23.5$. But it can be reduced down to a factor of $\sim 2$ if the normalization is made at fainter levels ($B = 18$), although in this case the observed bright count slope at $B < 18$ is steeper than predicted. For a discussion on this see[1].

In order to explain the discrepancy between model predictions and observations, a number of solutions concerning the somewhat *ad hoc* hypothesis on the absence of evolution in the galaxy population have been so far proposed. If galaxies were brighter in the past, for the same apparent magnitude range they would be observable at greater redshifts. In fact it has been shown that with a proper modelling of the galaxy luminosity evolution a nice fit to the counts down to very faint levels can be provided[1]. The main criticism to this approach comes from the absence of galaxies seen at high redshift in the surveys. To overcome this problem it has been suggested[4,5] that, together with some amount of luminosity evolution, there is also number density evolution. This is assumed to decrease on time, as galaxies merge to build up the present day population. Now, by reducing the amount of luminosity evolution, the expected number of galaxies at high redshift is also reduced. And so, the shape of the redshift distribution predicted from these *merging* models is quite similar to that from non-evolving.

*Present address: Instituto de Astrofisica de Andalucia, P.O. Box 3004, 18080 Granada, SPAIN

A basic ingredient to predict the number counts is of course the LF, which is normally fitted using catalogs of bright, nearby galaxies. The reliability of the LF at its faint end has been largely questioned[6], because intrinsically faint objects such as dwarfs or low surface brightness galaxies are not properly represented in those catalogs. In fact, it has been pointed out[7] that by accounting properly for the intrinsically faint population in the LF it could be possible to fit the counts even in the simple non-evolving case.

The study of the redshift distribution $N(z)$ of the faint field population could clearly give some insight onto which of these approaches (or combination of them) would be the most appropiate to fit the counts. The fact that different models provide of quite different $N(z)$ predictions makes it a crucial test to combine with the number counts.

## 2. The redshift distribution of galaxies at B=23-24 mag.

Spectroscopic observations of about $\sim 100$ galaxies selected in the B-band to have magnitudes B=23-24 have been reported up to now[8,9,10]. Even if the number of galaxies is rather small, some interesting results have already been achieved. Cowie et al. observed 12 galaxies in their survey, and could measure the redshifts of 11 of them. 4 out of the 11 objects turned out to be intrinsically faint galaxies located at relatively low redshifts. Glazebrook et al. and Campos et al. have reported observations of 55 and 53 galaxies respectively in that magnitude range. Unfortunately, the completeness rate of both samples is rather small, as it was only possible to measure the redshifts of about $\sim 63\%$ of the galaxies. The redshift distribution of the identified population (ID; i.e. galaxies with positive redshift identification) has a shape which is consistent with the predictions from non-evolving models, as shown in Figure 5 of[11]. This result points out that the *key* to solve the problem of the excess of faint field counts might be found in the redshift distribution of the unidentified population (un-ID). ID and un-ID galaxies do not only differ by the fact that the latters show featureless spectra in the optical window, but also by their colors. In Figure 1 it is plotted the B-R color distribution for the ID and un-ID populations in[9,10]. As it can be seen, un-ID galaxies are much bluer, on average, than IDs.

The fact that un-ID galaxies are bluer is, at first sight, a bit surprising. Blue colors are usually a sign of recent star formation activity in a galaxy. But the presence of strong emission lines in a spectrum is also associated with episodes of recent star formation, as they reveal the existence of ionizing photons from massive, young stars. Therefore it is expected that blue galaxies are those showing strong emission lines, and so the easiest to measure the redshifts. However it seems that, statistically, the redshifts of blue galaxies are the most difficult to identify.

A possible way to understand this apparent discrepancy would be by assuming that most of the un-ID objects are high redshift galaxies[10], such that the strongest emission lines are red-shifted outside the optical spectral range (as an example, [OII]Å3727 is red-shifted to $\sim 8000 Å$ at $z \sim 1$). A remaining problem is the absence of high redshift galaxies in the small, but rather complete sample of Cowie et al.

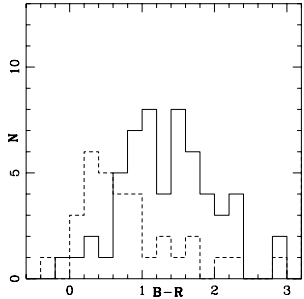

Fig. 1. B-R color histogram for ID (solid line) and un-ID galaxies (dashed line)

## 3. Dwarf Galaxies and the Faint Field Counts

As it was previously quoted[7], by accounting properly for the dwarf population in the LF it would be possible to fit the counts even by means of simple non-evolving models. The LF of dwarf galaxies (or, more general, of intrinsically faint galaxies) is very poorly known. Some recent work on this matter[11] has shown that, in nearby clusters, dwarfs are by far the dominant population. Drivers et al. used in their work a LF fitted for the dwarf population in clusters, normalized to fit the counts in the B-band. Even if this is a rather *ad hoc* LF, their work clearly illustrates that the rapid increase of counts at faint levels could be easily understood in terms of a, not unreasonable, large population of dwarfs, even without the neccesity of introducing any kind of strong evolution.

The redshift distribution of galaxies down to $B \sim 22$[12] has been found to be consistent with the predictions from non-evolving models. Normalizing the models at $B = 18$, the contribution from the faint end of the LF to the counts at those levels is expected to be small. Fainter than that, Cowie et al. detected an unexpected large population (for a *standard* LF) of intrinsically faint galaxies located at low redshifts. Further observational evidence on this could be found in the larger surveys of Glazebrook et al. and Campos et al., but unfortunately the incompleteness rate at the levels of interest was still rather large ($\sim 40\%$).

*Could the un-ID galaxies in these two surveys be dwarfs located at low redshifts?* It is now generally agreed that the star formation history of dwarf galaxies is different to that of spirals or irregulars. It has been suggested that, after a period of active star formation, the overlap of supernova explosions can develop a galactic wind blowing up most of the gas to the outerparts of these small systems with shallow potential wells, or even to the surrounding interstellar medium, from where could perhaps re-collapse given vise to new episodes of star-forming activity. Then, star formation would proceed in the form of strong, short *burts* of star formation followed by periods of quiescent activity, a view sustained by the low metallicities observed in these systems.

In Figure 2 it is shown the time evolution of the equivalent width of $[OII]\mathring{A}3727$ as well as the $B-R$ color after a single burst of star formation (the model has kindly been provided by G. Magris). The emission from the line is very strong during the first $\sim 10^6$ yrs. As massive stars evolve, the number of ionizing photons decreases very fast and so the equivalent widths of the emission lines. On the contrary, the color of the galaxies remains quite blue for a much longer period of time ($\sim 10^8 - 10^9$). Therefore, the simple association of blue color and strong emission lines would no longer be true, at least for dwarf galaxies.

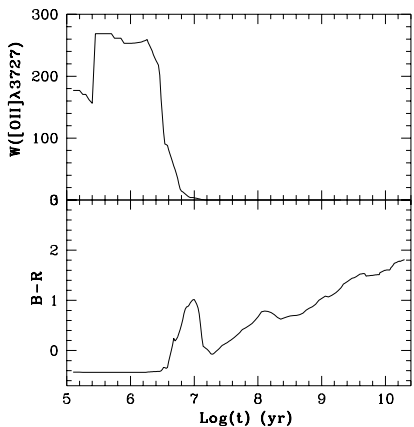

Fig. 2. Time evolution of the [OII]3727 equivalent width and the B-R color, for an instantaneus burst of star formation (model provided by G. Magris)

If star formation in dwarfs proceeds in a *bursting* mode, then the most common dwarf in the field, as seen in a B-selected sample, would be quite blue, but showing an otherwise featureless spectrum (as the galaxies become redder also become too faint to enter in a B-selected catalog). In general blue dwarf galaxies are thought to be characterized by having strong narrow emission lines superimposed to a rather weak, blue continuum. However it must be noticed that catalogs of blue dwarfs are in general built up by using as selection criteria either the existence of emission lines or strong UV excess. So that the catalogs are totally biased toward those galaxies experiencing an episode of star formation at present.

## 4. Summary and Conclusions

In order to account for the excess of galaxies observed in the blue-bands number counts with respect to the non-evolving predictions it is neccesary to assume some galaxy luminosity evolution and/or number density evolution, and/or to include in the LF a large population of intrinsically faint objects. The redshift surveys that have been undertaken to decide which of these model is actually at works, have failed to find a definitive answer due to the high rate of incompleteness achieved at faint magnitudes, where the excess starts to be important.

Un-ID galaxies at faint levels are charaterized by featureless spectra (this is of course the reason why the redshifts are not identified) and for being bluer, on average, than ID galaxies. This would support the view that most of the un-ID objects are high redshift galaxies, with the strongest spectral features red-shifted outside the optical window. The un-ID population would then be the high redshift tail predicted by pure luminosity evolution models.

However, the most complete redshift survey down to B=24 that exists at present[8] has failed to find any galaxy at very high redshift. On the contrary there appears to be an excess of intrinsically faint objects, located at relatively low redshifts. This result would then support the view that many of the blue galaxies seen in excess in the counts are intrinsically faint objects, and so to predict the counts we might include the existence of this population in the LF. Could the un-ID population be mainly built up by these dwarf objects? Assuming that the star formation in dwarfs proceeds in the form of single bursts followed by periods of quiescent star-forming activity, we have shown that the most common dwarf in the field, as seen in a blue-selected sample, would be quite blue, but showing a featureless spectrum. The two properties observed among the un-ID population.

Needless to say, the last word on this subject will not be written before we are able of identifying the redshifts of the un-ID population.

## 5. References

1. N. Metcalfe, T. Shanks, R. Fong and N. Roche, *MNRAS* (1994) in press.
2. J.A. Tyson, *Astron. J.* **96** (1988) 1.
3. S.J. Lilly, L.L. Cowie and J.P. Gardner *Astrophys. J.* **369** (1991) 79.
4. B. Guidernoni and B. Rocca-Volmerange, *Astron. Astrophys.* **252** (1991) 435.
5. T.J. Broadhurst, R.S. Ellis and K. Glazebrook *Nature* **355** (1992) 55.
6. S. Phillips and T. Shanks *MNRAS* **227** (1987) 115.
7. S.P. Drivers, S. Phillips, J.I. Davies, I. Morgan and M.J. Disney *MNRAS* **266** (1994) 155.
8. L.L. Cowie, A. Songalia and E.M. Hu *Nature* **354** (1991) 460.
9. K. Glazebrook, R. Ellis, M. Colless, T. Broadhurst, J. Allington-Smith, N.R. Tanvir and K. Taylor (1994), preprint.
10. A. Campos, T. Shanks, N. Metcalfe, N. Roche and N.R. Tanvir, in preparation.
11. B. Binggeli, A. Sandage and E.A. Tamman *Astron. J.* **90** (1985) 1681.
12. M.M. Colless, R.S. Ellis, K. Taylor and R.N. Hook *MNRAS* **244** (1990) 408.